# On the role of secondary electron emission in capacitively coupled radio frequency plasma sheath: a theoretical ground

Guang-Yu Sun, Han-Wei Li, An-Bang Sun*, Yuan Li, Bai-Peng Song, Hai-Bao Mu, Xiao-Ran Li and Guan-Jun Zhang†

_________

Research Center for Advanced High Voltage and Plasma Technology, State Key Laboratory of Electrical Insulation and Power Equipment, School of Electrical Engineering, Xi'an Jiaotong University, Xi'an, Shaanxi, 710049, China

† gjzhang@xjtu.edu.cn
* anbang.sun@xjtu.edu.cn

_________

We propose a theoretical ground for emissive capacitively coupled radio-frequency plasma sheath under low pressure. The rf sheath is assumed to be collisionless, and oscillates with external source. A known sinusoidal voltage instead of current is taken as prerequisite to derive sheath dynamics. Kinetic studies are performed to determine mean wall potential as a function of secondary emission coefficient and applied voltage amplitude, with which the complete mean DC sheath is resolved. Analytical analyses under homogeneous model and numerical analyses under inhomogeneous model are conducted to deduce real-time sheath properties including space potential, sheath capacitance and stochastic heating. Obtained results are validated by a continuum kinetic simulation without ionization. The influences of collisionality and ionization induced by secondary electrons are elucidated with a particle-in-cell simulation, which further formalizes proposed theories and inspires future works.

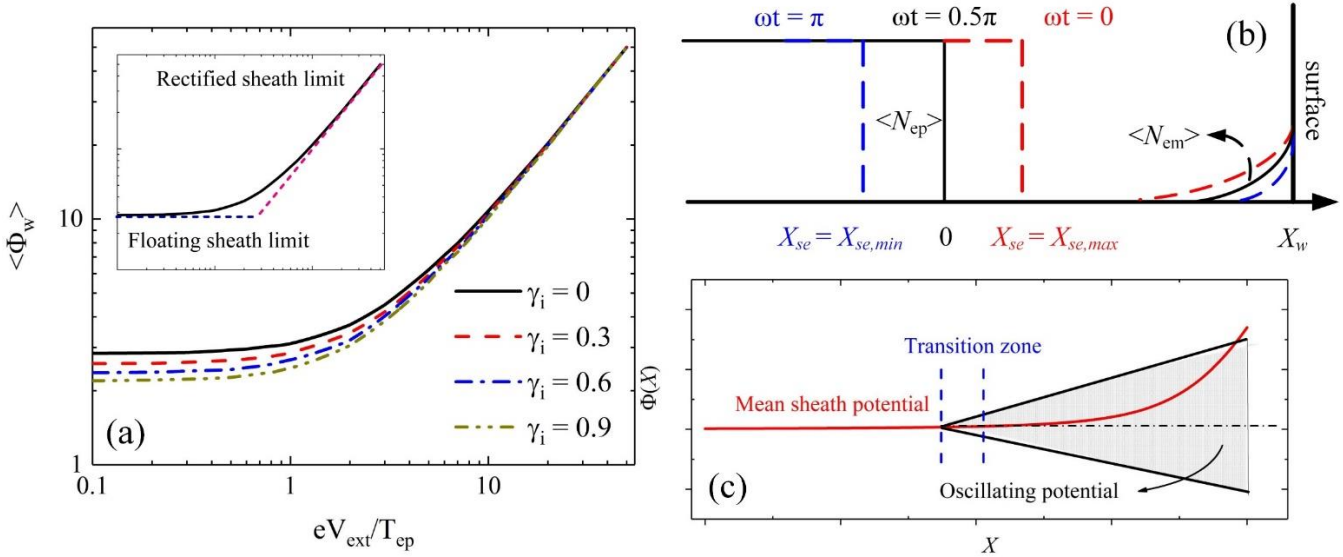

# 1 Introduction

Low-pressure capacitively coupled plasma (CCP) has been widely used in a variety of plasma processing applications in semiconductor industry, such as etching, deposition and many other surface treatment technics.[1-5] CCP is commonly generated by applying radio-frequency (rf) voltage or current to electrodes immersed in plasma, which creates high voltage capacitive sheath between the electrode and plasma bulk. Boundary effects, namely plasma-surface interactions (PSI), are of vital significance in understanding capacitively coupled radio-frequency (CCRF) sheath,[6-8] among which secondary electron emission (SEE) induced by ion flux is one of the most common features and is worth investigating.[6, 7] The aim of this paper is to establish an *ab initial* theoretical ground on rf sheath of electron-emitting surface, aka emissive rf sheath, and facilitate future works regarding CCP discharge in either concept or application.

Modeling capacitive discharge self-consistently is complicated even in the simplest 1D planar configuration. Therefore, many assumptions were employed in previous theoretical works to obtain analytical solutions. Early attempts of Godyak and Sternberg,[9] as well as Lieberman[10] constructed CCP model with known sinusoidal current and electron density in the form of step function. Kaganovich performed detailed kinetic studies in low-pressure rf discharge and specially considered rf field penetration using two-step ion density assumption.[11, 12] Recent study of Turner and Chabert introduced a new ansatz to enable analyses of complex waveform, which postulates that mean electron density is a constant fraction of ion density.[13] Above models choose a sinusoidal (or sum of a series of sinusoidal) current(s) as the prerequisite to deduce discharge parameters such as sheath potential, power transfer, motion of electron sheath front, etc. In experiment or particle-in-cell (PIC) simulation, a known sinusoidal voltage is more frequently adopted. Analytical model proposed by Riemann,[14] as well as Czarnetzki,[15] performed alternative deductions based on applied rf voltage. Fluid model given by Heil *et al* also implemented known rf wall potential.[16] These models significantly advance the understanding of CCRF sheath.

In practical CCP applications, the boundary is not always perfect absorber as many of above models assumed. Secondary electron emission (SEE) on solid surface induced by either electron or ion is ubiquitous in many plasma apparatuses like Hall thruster, Langmuir probe and dielectric barrier discharge, etc.[17-20] In terms of CCP discharge between two metallic electrodes, ion-induced SEE is more predominant and extensive amount of experimental and simulation works have been conducted to illustrate its influences. Early work of Misium *et al* involved SEE in their rf discharge model using fluid theory with constant secondary electron (SE) velocity.[21] Lafleur *et al* proposed a concise expression to characterize electrical asymmetric effect induced by different secondary emission yield (SEY) $\gamma_i$, also called emission coefficient.[22] A series of simulations were performed which incorporated ion-induced secondary electron emission (ISEE) and illustrated its influences on discharge parameters,[7, 23, 24] showing some differences with simulations where ISEE is not involved.[25]

Though it has been illustrated that the presence of SEE modifies rf sheath dynamics, the exact relations between sheath potential, capacitance, conductance and different emission coefficients/applied voltages have not been clarified, on account that in simulation most discharge parameters are closely coupled. Theoretical deduction based on plasma kinetic theory, as an alternative approach, is able to determine the role of secondary electrons in CCP self-consistently and provide resolute plasma sheath solution, which is the main purpose of this work. Presented model is justified by numerical calculation as well as two distinct simulations designed in collisionless and weakly collisional regime, respectively.

In section 2.1 we perform kinetic studies to derive mean sheath potential by determining mean wall potential and solving Poisson equation as an initial value problem. Real-time sheath potential and sheath capacitance are analyzed using analytical approaches with homogeneous model, as well as numerical approaches with inhomogeneous model in sections 2.2 and 2.3, respectively. In section 3 obtained results are compared to a continuum, kinetic simulation, which is found to be consistent with deduced theories. Some discrepancies due to model assumptions are elucidated as well. Influences brought by higher collisionality are expatiated with a particle-in-cell simulation code. Concluding remarks are given in section 4.

## 2 Theoretical Modeling of Emissive RF Sheath

To investigate the intricate sheath oscillation in response to external source, a more expedient method is to first calculate mean DC sheath and then tackle the time-dependent sheath. The presence of secondary electrons modifies particle as well as energy flux in both floating and rf sheath, while the key difference between these two conditions is that zero-current condition is not necessarily satisfied at all time with respect to a rf sheath. As a matter of fact, change of flux balance due to SEE in rf sheath is important only in a time-averaged manner. In addition, since conduction current of electrons in sheath is negligible in contrast to displacement current, SE's density should be small compared with plasma density and is unlikely to produce remarkable impact in terms of time-dependent space potential. Consequently, one can imagine that SEE plays different roles in mean DC sheath and time-dependent sheath. In the following discussions, we will first derive potential of the mean DC sheath and then calculate real-time space potential using two distinct models.

### 2.1 Time-averaged emissive rf sheath

To calculate the mean DC sheath, we will make use of the flux balance and electron velocity distribution function (EVDF) of SEs to express mean wall potential with emission coefficient as well as amplitude of applied voltage. Poisson's equation is to be transformed into an initial value problem (IVP), where calculated wall potential serves as the initial condition. Some basic assumptions and adopted notations are provided below before further deductions.

Temporal potential variation in rf sheath is consist of a time-averaged component in space plus an oscillating component, expressed in the following form:

$$V(x,t) = \overline{V(x)} + V_{osc}(x,t) \qquad 2.1.1$$

where $V(x,t)$ is the spatialtemporal potential, $\overline{V(x)} = \langle V(x,t) \rangle_t$ is the mean DC sheath potential, and the mean value of oscillating component $\overline{V_{osc}(x,t)} = 0$. Particularly, when a sinusoidal external voltage source is supplied, potential at boundary becomes:

$$V_w(x_w,t) = \overline{V_w} + V_{ext}\cos(\omega t) \qquad 2.1.2$$

where $x_w$ is the position of boundary, $\overline{V_w}$ is the mean potential at wall $\overline{V(x_w)}$, and $V_{ext}$ (which is positive) is amplitude of voltage source between electrode and plasma bulk with angular frequency $\omega$. The location of electron sheath edge at $\omega t = \frac{\pi}{2}$ is set as ion sheath front and equals to 0, on which external source vanishes. Detailed definition of sheath boundary adopted in calculation is to be dwelled on later. Note that ion front is static since ions respond only to time-averaged potential, while electron front (instantaneous sheath edge) moves away from the wall when $\omega t \in [0, \pi]$. To solve for the exact space potential, its boundary condition namely $\overline{V_w}$ has to be derived in the first place.

In the following context, we implement plasma kinetic theory to determined $\overline{V_w}$ based on discharge parameters. Several normalized values are introduced below to facilitate deductions:

$$\Phi_s = -\frac{eV_s}{T_{ep}}, N_s = \frac{n_s}{n_0}, \Theta = \frac{T_{ep}}{T_{em}}, \Xi = \frac{\varepsilon_{i0}}{T_{ep}}, \mu = \frac{m_i}{m_e}, X = \frac{x}{\lambda_{De}} \qquad 2.1.3$$

Here $T_{ep}$ is plasma electron temperature, $T_{em}$ is temperature of secondary electrons from surface, $n_0$ is plasma density at static sheath edge (not to be confused with density in plasma bulk), $\varepsilon_{i0}$ is initial ion energy at sheath edge, $m_i$ and $m_e$ are masses of ion and electron, respectively, $\lambda_{De}$ is electron Debye length at sheath edge. The subscript $s$ can be $w$, $ext$, $ep$, $em$, or $i$ representing wall, external source, plasma electron, secondary electron and ion.

Plasma electron flux at wall is dictated by Hertz-Langmuir formula if loss cone is ignored:

$$\Gamma_{ep}(\Phi_w) = \frac{1}{4} n_{ep0} \sqrt{\frac{8T_{ep}}{\pi m_e}} \exp(\Phi_w) \qquad\qquad 2.1.4$$

with $\Phi_w = -\frac{e\overline{V_w}}{T_{ep}}$ and $n_{ep0}$ is plasma electron density at sheath edge. It is worth noting that the loss cone exists because energetic electrons that penetrate sheath potential cannot return to plasma, thus EVDF depletes at high velocity. Recent study confirmed that:[26]

$$\frac{\Gamma_{ep}}{\Gamma_{ep,loss\,cone}}\bigg|_{wall} = \frac{1 + \mathrm{erf}(\sqrt{\Phi_w})}{2} \qquad\qquad 2.1.5$$

Notably, the effect of loss cone is more obvious for floating sheath where $\Phi_w$ is small compared with high voltage sheath. In most CCP discharge, the strong external voltage leaves few electrons crossing the sheath. In this regard, involving equation 2.1.5 can hardly provide better precision though it complicates the calculation. Thus, equation 2.1.4 is adopted for a clearer expression which assumes Boltzmann distribution of plasma electrons.

The initial ion flux from presheath is usually prescribed by the well-known Bohm criterion, yet it is necessary to discuss its validity in emissive rf sheath. *Ab initio* analyses given by Meijer suggested a weaker condition for a rf sheath to exist, with $\Xi \geq 0.5(1 + \overline{(\frac{V_{osc}}{\overline{V}})^2})^{-1}$,[27] while Riemann considered the electrostatic wave entering sheath edge and confirmed that Bohm criterion is still valid if instantaneous field is counted.[28] Regarding electron emission from boundary, the Bohm criterion is found to be barely modified with weak electron-induced secondary electron emission (ESEE).[29, 30] In addition, our recent work showed that the Bohm criterion holds for ion-induced electron emission by applying its general form at marginal condition.[26] With all these, we are safe to continue with $\Xi = 0.5$. Since ions are inertial and only respond to the time-averaged field, ion flux in collisionless limit is constant and is expressed by:

$$\Gamma_i = n_0 \sqrt{\frac{T_{ep}}{m_i}} \qquad\qquad 2.\,1.\,6$$

A key difference between emissive rf sheath and sheath near perfectly-absorbing boundary is the presence of secondary electron beam. In a multitude of previous models, a common assumption is that electron density profile is described by step function as sheath size $s \gg \lambda_D$, where region out of electron sheath front contains no electron for a typical high voltage sheath.[9, 10, 14] A contradiction, however, arises when SEE is included because density of secondary electrons is most intense where least plasma electrons present. Additionally, SEs are extensively accelerated by the high voltage sheath which replenishes the high velocity end in EVDF of bulk plasma, making the exact behavior of SEs rather intricate. Thus it is necessary to clarify which effect brought by SEE is most determinant to derive a resolute sheath solution within achievable complexity. In order to calculate the potential at surface, we will first determine the time-averaged distribution of secondary electrons.

Secondary electrons from boundary can be described by a half-Maxwellian $f_{em}$ with temperature $T_{em}$, whose value is usually lower than plasma electron temperature (typically 1~3 eV).[31, 32] Flux of SEs is easily obtained with $\Gamma_{em}|_{wall} = \int_0^{+\infty} f_{em} v \mathrm{d}v$, which gives:

$$\Gamma_{em}(\Phi_w) = N_{em}(\Phi_w) \sqrt{\frac{2T_{em}}{\pi m_e}} \qquad\qquad 2.\,1.\,7$$

Both kinetic theory and fluid model can be applied to study SE's dynamics. We shall begin with kinetic analyses with specific distribution functions for secondary electrons, the fluid model will then be proved to be a limiting case of kinetic model when $T_{em} \to 0$. Density of SEs in collisionless sheath is obtained by integrating EVDF from lower bound $v_{min}(\Phi) = \sqrt{2T_{ep}(\Phi_w - \Phi)/m_e}$, leading to:

$$N_{em}(\Phi) = N_{em}(\Phi_w)\exp[\theta(\Phi_w - \Phi)]\mathrm{erfc}[\sqrt{\theta(\Phi_w - \Phi)}] \qquad\qquad 2.\,1.\,8$$

Note that quantities in above equation are time-dependent because space potential oscillates over time, with $\Phi_w(t) = -\frac{e}{T_{ep}}[\overline{V_w} + V_{ext}\cos(\omega t)]$. This indicates that SE's density varies over time but equation 2. 1. 8 is always valid as $\omega \ll \omega_e$ and electrons respond instantaneously to change of space potential.

With above deductions, the average wall potential $\overline{\Phi_w} = -\frac{e}{T_{ep}}\overline{V_w}$ can be derived from flux balance and charge neutrality. Following two equations represent secondary electron emission from electrode and charge neutrality at sheath edge, respectively:

$$\gamma_i \Gamma_i = \Gamma_{em} \qquad\qquad 2.1.9$$

$$N_{em}(0) + N_{ep}(0) = 1 \qquad\qquad 2.1.10$$

where $\gamma_i$ is the secondary emission yield due to ion bombardment. Since there is no net conduction current in sheath, total wall influx including electrons and ions must be zero in an average sense (but not instantaneously):

$$\overline{\Gamma_{net}} = \overline{\Gamma_i - (\Gamma_{ep} - \Gamma_{em})} = 0 \qquad\qquad 2.1.11$$

Combining above equations, we arrive at:

$$(1 + \gamma_i)\sqrt{\frac{2\pi}{\mu}} = \left\langle \left[1 - \gamma_i\sqrt{\frac{\pi\Theta}{2\mu}}\exp(\Theta\Phi_w)\mathrm{erfc}(\sqrt{\Theta\Phi_w})\right]\exp(-\Phi_w)\right\rangle_{2\pi} \qquad 2.1.12$$

where RHS of above equation is averaged in one period of time. One can justify derived equation with two limiting cases. When $\gamma_i = 0$, equation 2. 1. 12 is reduced and an analytical solution is obtained as follows:

$$\overline{\Phi_w}\big|_{\gamma_i=0} = \ln\left(\sqrt{\frac{\mu}{2\pi}}\right) + \ln[I_0(-\Phi_{ext})] \qquad\qquad 2.1.13$$

where $I_0$ is the modified Bessel function of the first kind in order zero and $\Phi_{ext} = -\frac{eV_{ext}}{T_{ep}}$. Naturally, the first term in RHS is the floating sheath potential, and the second term in RHS is determined by external voltage which equals to zero if $\Phi_{ext} = 0$. It is clear that average sheath potential increases with applied voltage amplitude. From the asymptotic expansion it is found that at high voltage limit $\lim_{V_{ext}\to+\infty} \frac{\overline{\Phi_w}}{\Phi_{ext}} = 1$ and $\left|\frac{\overline{\Phi_w}}{\Phi_{ext}}\right| > 1$. Therefore, plasma potential always floats well above the wall, which is also correct considering ion-induced SEE. Condition can be somewhat different at higher emission rate.[33]

Calculation results of non-emissive (i.e. $\gamma_i = 0$) sheath potential are given in black curve of figure 1(a).The two extremities behave as they should. When $V_{ext}$ is small the sheath is close to a floating sheath, which transforms into a rectified sheath with large $V_{ext}$.

A different method to derive equation 2. 1. 8 is to directly apply the flux and energy conservation with SE's initial velocity being set as $v_{em0} = \sqrt{\frac{2T_{em}}{\pi m_e}}$. Obtained expression of SE's density in sheath is:

$$N_{em}(\Phi) = N_{em}(\Phi_w) \frac{v_{em0}}{\sqrt{v_{em0}^2 + 2T_{ep}(\Phi_w-\Phi)/m_e}} \qquad 2.\ 1.\ 14$$

When $T_{em} \ll T_{ep}$, it gives $\lim_{T_{em}\to 0} \frac{N_{em}(\Phi)}{N_{em}(\Phi_w)} = \frac{1}{\sqrt{\pi\theta(\Phi_w-\Phi)}}$, leading to a simplified form of SE's density:

$$N_{em}(\Phi) = \frac{N_{em}(\Phi_w)}{\sqrt{\pi\theta(\Phi_w-\Phi)}} \qquad 2.\ 1.\ 15$$

Equation 2. 1. 8 converges into the same result when applying $\theta \to \infty$, which produces a singularity at $\Phi = \Phi_w$. Such approximation simplifies equation 2. 1. 12 and gives the following expression:

$$\overline{\Phi_w} = \ln\left(\frac{1}{1+\gamma_i}\sqrt{\frac{\mu}{2\pi}}\right) + \ln[I_0(-\Phi_{ext}) - \frac{\gamma_i}{2\pi\sqrt{2\mu}}H] \qquad 2.\ 1.\ 16$$

with the definite integral defined as $H = \int_0^{2\pi} (\overline{\Phi_w})^{-0.5} \exp[-\Phi_{ext}\cos(\omega t)]\mathrm{d}(\omega t)$. Note that the first term in RHS is the floating sheath potential considering ion-induced SEE with $\Phi_{w,float} = \ln(\sqrt{\frac{\mu}{2\pi}}\frac{1}{1+\gamma_i})$, deduction can be found in our recent work.[26]

Numerical solution of equation 2. 1. 12 is given in figure 1(a)(b). Note that the range of $\gamma_i$ is usually within 0.5[5,22,26], higher values are adopted in some cases to make the trend more obvious and enhance readability. It can be seen that the general trend between two limits are analogous, where average sheath potential decreases with emission yield $\gamma_i$. The reduction of wall potential due to SEE can also be found in equation 2.1.16 as the first term in RHS decreases with $\gamma_i$ while the second term is very small when $eV_{ext} \ll T_{ep}$. Additionally, in figure 1(b)(c) it is shown that equation 2.1.12 and equation 2.1.16, or in other word, using equation 2.1.8 or equation 2.1.15 as SE's density, provide almost the same results. Approximation of SE's distribution given by equation 2.1.15 will be used again later on.

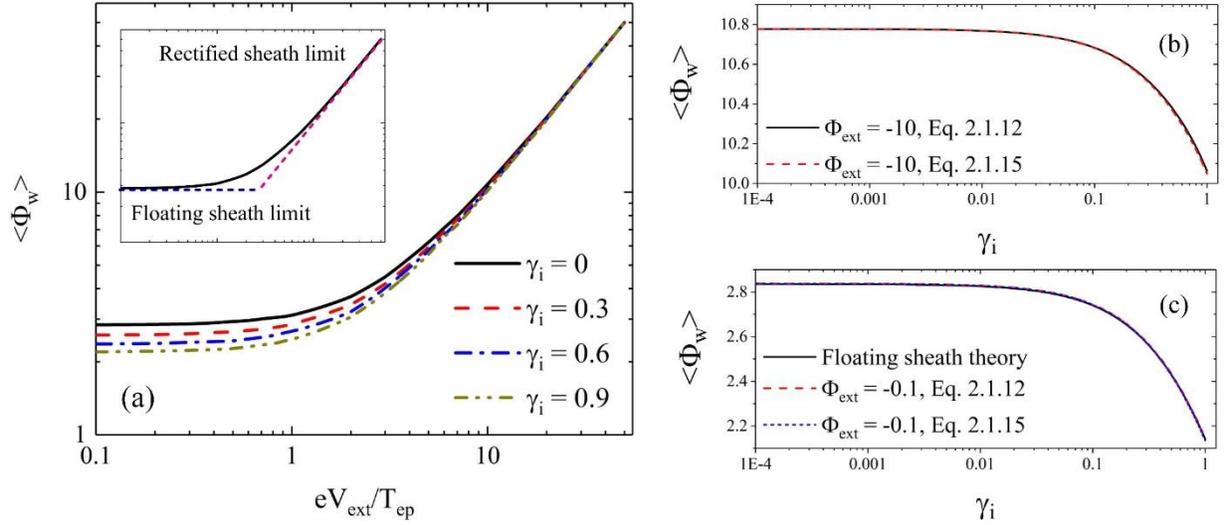

Figure 1. Calculation results of mean wall potential with different $\Phi_{ext}$ and $\gamma_i$. Two limiting cases are shown in (a), and two expressions of SE's density are compared in (b), (c). Note that floating sheath in (c) is calculated with $\Phi_{w,float} = \ln(\sqrt{\frac{\mu}{2\pi}}\frac{1}{1+\gamma_i})$.

Above deductions relate wall potential to applied voltage and emission yield. In the following context, obtained mean wall potential is used as boundary condition to help solve mean DC sheath in space.

Starting from Poisson's equation and ion fluid model, space potential is prescribed by the following expressions:

$$\frac{d^2\overline{\Phi}}{dX^2} = N_i(\overline{\Phi}) - \overline{N_e}(\Phi) \qquad 2.1.17$$

$$N_i(\overline{\Phi}) = (1 + 2\overline{\Phi})^{-0.5} \qquad 2.1.18$$

Remarkably, average electron density $\overline{N_e}(\Phi)$ is simplified as $N_e(\overline{\Phi})$ if oscillation is not significant.[14] Average electron density contains plasma electron density as well as SE's density, while the former follows Boltzmann distribution and the latter is dictated by equation 2.1.15. To solve equation 2.1.17, multiply it with $\frac{d\overline{\Phi}}{dX}$ and integral from 0 to $\overline{\Phi}$, we arrive at:

$$\frac{1}{2}(\frac{d\overline{\Phi}}{dX})^2 = \left(\sqrt{1+2\overline{\Phi}}-1\right) - \left(1 - \frac{\gamma_i}{\sqrt{2\mu\overline{\Phi_w}}}\right)[1-\exp(-\overline{\Phi})] - \gamma_i\sqrt{\frac{2}{\mu}}(\sqrt{\overline{\Phi_w}} - \sqrt{\overline{\Phi_w} - \overline{\Phi}}) \qquad 2.1.19$$

Equation 2.1.19 can be viewed as an initial value problem with initial condition $\overline{\Phi}(X_w) = \overline{\Phi}\big|_{wall} = \overline{\Phi_w}$, and space potential is derived by solving IVP from the wall towards the sheath edge. The exact location of sheath edge viewed from the sheath scale is $-\infty$ relative to the wall according to previous two-scale analyses of sheath structure by Riemann.[34] Here to give numerical solution, a sufficiently large interval ($20\lambda_{De}$) is chosen and the location of sheath edge is determined where electron density diverges from that of ion by $\delta_{se} = \frac{n_i - n_e}{ni} = 5\%$. Notably, the location of static sheath edge is not decided from Bohm criterion but is based on a more practical concept here. This approach, i.e. choosing a significant breakdown of charge neutrality, was originally adopted by Tonks and Langmuir,[35] while recent study of Chabert confirmed that such choice makes DC sheath size nearly independent of plasma properties,[36] thus facilitating comparison under various conditions. A value of $\delta_{se} = 1\%$ was chosen in Campanell's recent work.[37] The influence of such choice on sheath size is shown in figure 2. Clearly the scale of sheath drops less sharply after 5%. It also varies accordingly with $\gamma_i$ and $\overline{\Phi_w}$. In the context of this article, such value only modifies the location as well as potential of static sheath edge, which does not bring essential changes in CCP discharge.

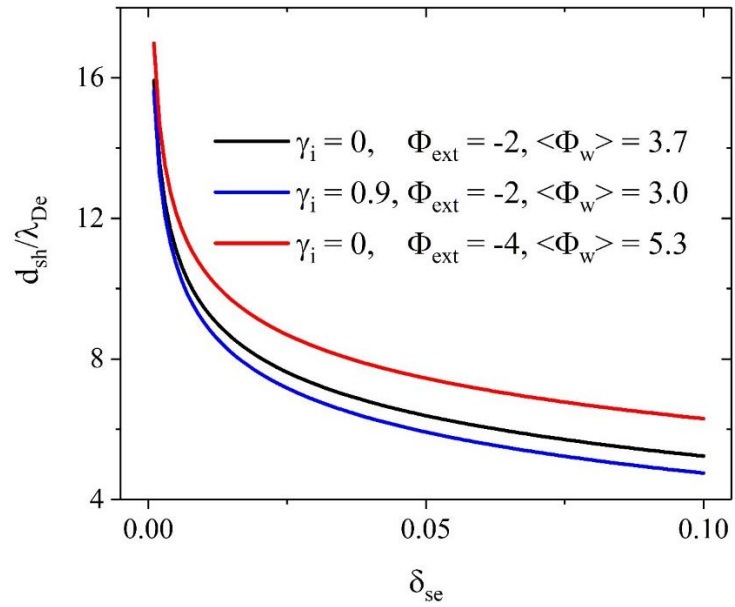

Figure 2. Obtained sheath size with different choices of $\delta_{se}$, emission coefficient and external voltage. Note that $d_{sh}$ is the size of mean DC sheath.

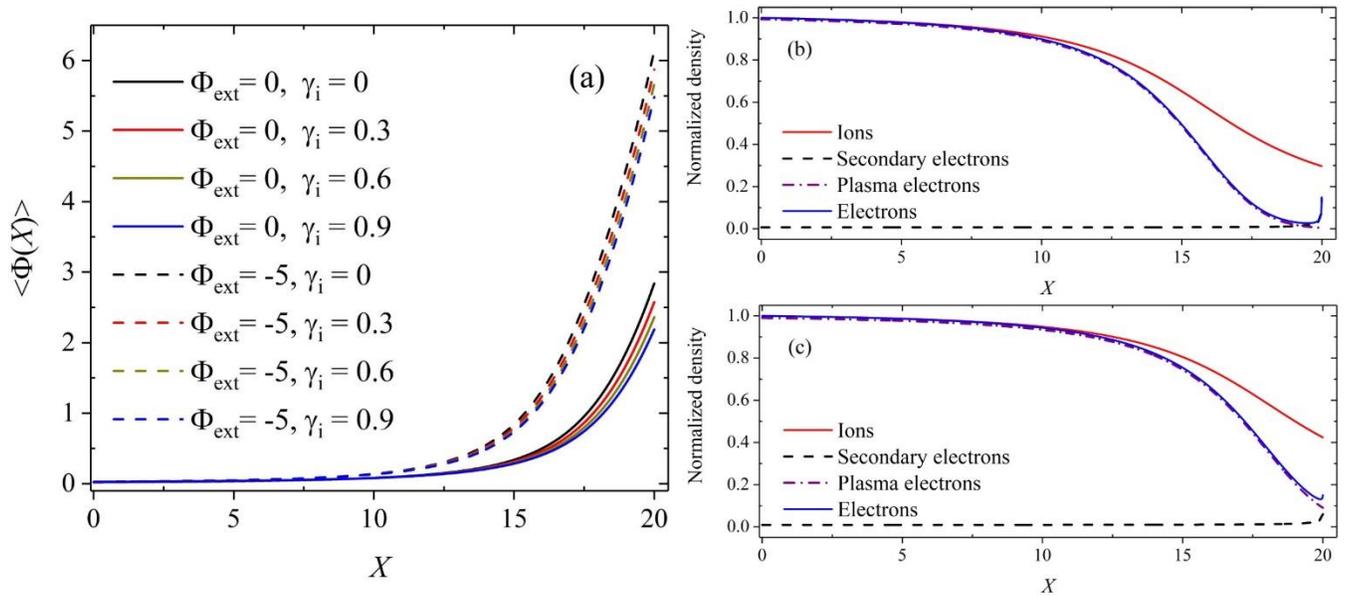

Figure 3. (a) Calculated mean DC potential in space, with different external voltages and emission coefficients. (b) Charge density in sheath using equation 2. 1. 18, a singularity appears near surface. (c) Charge density in sheath using equation 2. 1. 8, SE's density is smaller than panel (b). Note that in (b) and (c) $\Phi_{ext} = 5$ and $\gamma_i = 0.9$. A high value of emission coefficient is chosen to enhance visibility.

Obtained mean DC sheath potentials are shown in figure 3. Space potential presented in figure 3(a) is consistent with previous results that potential at surface is mitigated by SEE. Figure 3(b) indicates that the presence of SEE makes electron density profile non-monotonic and modifies ion density distribution. In addition, secondary electron density quickly declines away from the boundary while plasma electron density almost vanishes adjacent to the wall due to intense potential barrier. However, results in figure 3(b) somewhat exaggerates the density of secondary electrons and produces an infinite SE density on the surface. A more precise expression of $N_{em}$ is applied in figure 3(c) where one can find that the sharp rise of SE's density near surface disappears and less SEs exist on boundary.

In above deductions, one may question that why a seemingly tiny amount of secondary electrons which primarily concentrate near surface could remarkably modify the space potential in entire sheath? To answer this, it is necessary to understand that the change of emission yield mainly manages the mean wall potential $\overline{\Phi_w}$ by virtue of floating condition (i.e. flux balance), which serves as the initial condition in equation 2. 1. 19. Once the mean wall potential is fixed, the presence of SEs has few effects on space potential. This can be quickly justified by setting same $\overline{\Phi_w}$ calculated from non-zero $\gamma_i$ and forcing $\gamma_i = 0$ in equation 2. 1. 19. Calculation result is almost the same as that using the non-zero $\gamma_i$, see figure 4. This property is important when resolving temporal space potential in the next section.

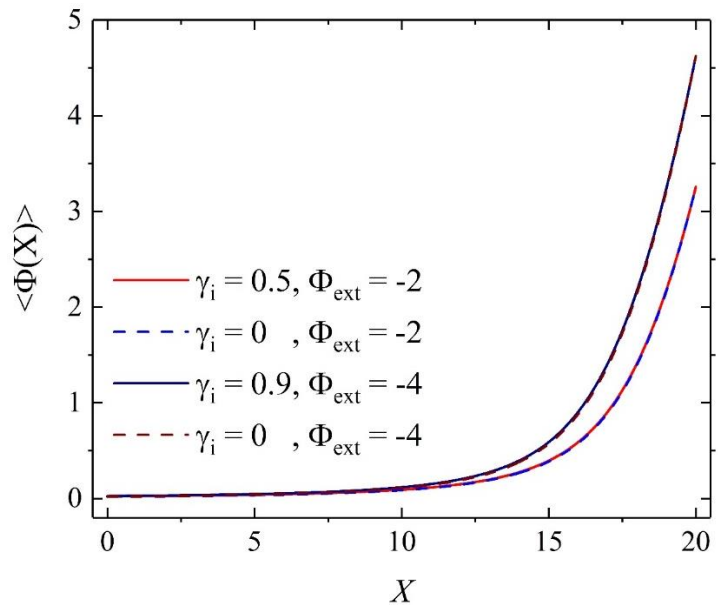

Figure 4. Comparison of calculated mean sheath potential. Wall potentials are obtained and fixed on given parameters in first and third curve, which are used as initial condition following equation 2. 1. 19. In the second and fourth curve $\gamma_i$ is set as 0 when resolving equation 2. 1. 19. Clearly. once the wall potential is determined, SEE has few influences on space potential.

## 2.2 Analytical description of real-time emissive rf sheath with homogeneous model

In section 2.1, we have obtained the mean DC sheath. To derive the time-dependent sheath, an intuitive approach is to directly solve Poisson's equation with a static ion distribution and plasma electrons following Boltzmann distribution. However, such differential equation cannot be directly solved because the implicit relation between $\overline{\Phi}$ and $\Phi_{ocs}$ is unknown, making the reduction of order impossible like in equation 2. 1. 17, 2. 1. 19. Since the exact form of sheath potential at arbitrary moment is intricate to calculate, additional assumptions are in need to facilitate deductions. One common postulate is to replace the electron density drop in sheath with a step function, which enables us to trace the motion of electron sheath front and solve the sheath.

One major difference between homogeneous and inhomogeneous models lies in the assumption of ion density. Homogenous model postulates that ion density is constant in sheath while inhomogeneous model considers the drop of ion density in space. Though the former has several drawbacks (which we will show later), it yields analytical expression to help understand the properties of an emissive rf sheath in an explicit manner. In this section, we will first introduce this homogeneous model and then clarify how SEE influences sheath dynamics.

To give real-time evolution of space charge, it is necessary to first derive the oscillating component in equation 2. 1. 1. An intuitive method is to rewrite Poisson equation while separating the two potential terms, which leads to the equations below:

$$\frac{\mathrm{d}^2\overline{\Phi}}{\mathrm{d}X^2} = N_i(\overline{\Phi}) - N_e(\overline{\Phi}) \qquad\qquad 2.\ 2.\ 1$$

$$\frac{\partial^2\Phi}{\partial X^2} = N_i(\overline{\Phi}) - N_e(\Phi) \qquad\qquad 2.\ 2.\ 2$$

Note that $\langle N_e(\Phi)\rangle = N_e(\overline{\Phi})$ is used in deriving equation 2. 2. 1-2 by averaging real-time Poisson's equation. Similar treatment was also implemented in previous studies of RF sheath theory. [3,14] Bringing equation 2. 2. 1 into equation 2. 1. 2, we arrive at:

$$\frac{\partial^2\Phi_{osc}}{\partial X^2} = \left[N_{ep}(\overline{\Phi}) - N_{ep}(\Phi)\right] + \left[N_{em}(\overline{\Phi}) - N_{em}(\Phi)\right] \qquad\qquad 2.\ 2.\ 3$$

The two terms in RHS are to be solved separately in the following context. Using step function ansatz where electron density vanishes away from time-dependent density front $X_{se}(t)$, the first term in RHS of equation 2. 2. 3 can be derived. A schematic is shown in figure 5 containing the simplified electron density profile. We choose $X = 0$ at $\omega t = 0.5\pi$, on which space potential equals to mean DC sheath potential since no external voltage is applied.

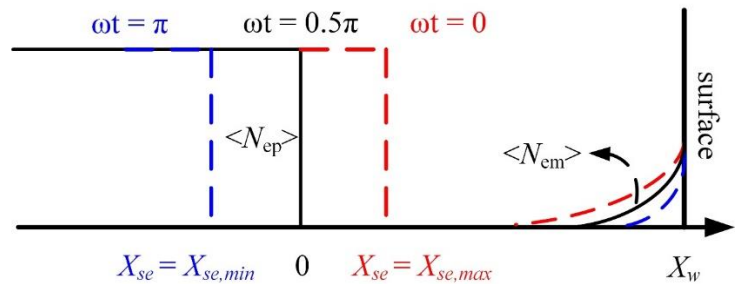

Figure 5. Schematic of plasma electron density.

Integrating equation 2.2.3 twice over $X$, one can derive the time-dependent sheath. Obtained results are given below:

$$\Phi_{osc} = \begin{cases} \Delta I_{em}(x,t), X \leq X_{LB} \\ -\frac{1}{2}N_{ep}(0)(X-X_{LB})^2 \text{sgn}[X_{se}(t)] + \Delta I_{em}(X,t), X_{LB} < X \leq X_{UB} \\ N_{ep}(0)\left[\frac{X_{se}^2(t)}{2} - X_{se}(t)X\right] + \Delta I_{em}(X,t), X > X_{UB} \end{cases} \qquad 2.2.4$$

$$\Delta I_{em}(X,t) = \int_0^X \mathrm{d}X'' \int_0^{X''} \left[N_{em}(\overline{\Phi}) - N_{em}(\Phi)\right] \mathrm{d}X' \qquad 2.2.5$$

with $X_{LB} = \text{Min}\{0, X_{se}(t)\}$ and $X_{UB} = \text{Max}\{0, X_{se}(t)\}$. Location of real-time plasma electron front is determined from boundary condition $\Phi_{osc}(X_w) = \Phi_{ext}\cos(\omega t)$, which yields $X_{se}(t) = X_w - \sqrt{X_w^2 - \frac{2}{N_{ep}(0)}[\Delta I_{em}(X_w,t) - \Phi_{ext}\cos(\omega t)]}$. Note that $X_w$ is the location of surface. If we define the normalized mean DC sheath size as $D_{sh} = \frac{d_{sh}}{\lambda_{De}}$ with $d_{sh}$ the size of mean DC sheath, the motion of sheath front is given by:

$$X_{se}(t) = D_{sh} - \sqrt{D_{sh}^2 - \frac{2}{N_{ep}(0)}[\Delta I_{em}(X_w,t) - \Phi_{ext}\cos(\omega t)]} \qquad 2.2.6$$

Equation 2.2.5 contains the unknown term $\Phi_{osc}$ so a direct solution would be intricate. However, the calculation can be greatly simplified by the fact that the length of transition zone $[X_{LB}, X_{UB}]$ is small(i.e. $|X_{se}(t)| \ll X_w$) and space potential varies quasi-linearly near the surface (not the case within $[X_{LB}, X_{UB}]$). Note that such approximation is valid since $N_{em}$ quickly drops down away from surface as one can clearly see in figure 3(b)(c). Equation 2.2.5 is simplified as follows:

$$\Delta I_{em}(X,t) = \int_0^X \mathrm{d}X'' \int_0^{X''} \left[N_{em}(\overline{\Phi}) - N_{em}\left(\overline{\Phi} + \frac{X'}{X_w}\Phi_{ext}\cos(\omega t)\right)\right] \mathrm{d}X' \qquad 2.2.7$$

where density of secondary electrons is given by equation 2.1.8 or equation 2.1.15. Calculated sheath potentials using above equations are shown in figure 6. It is worth noting that electron density given by equation 2.1.8 is not valid if the sheath potential is non-monotonic. For instance, if $\Phi(X)$ first decreases then increases, instead of monotonically rising towards the surface, there exists a minimum value within the sheath. Such potential distribution appears in analytical model when $\omega t \in (0, 0.5\pi)$.

Rigorously speaking, electron density on the left side of minimum voltage $\Phi_{min}(X_{min})$ should be $N_{em}(\Phi) = N_{em}(\Phi_w)\exp[\theta(\Phi_w - \Phi)]\mathrm{erfc}[\sqrt{\theta(\Phi_w - \Phi_{min})}]$, which is implemented when calculating results in figure 6. However, this calibration brings little changes in value as density of secondary electrons quickly shrinks away from the boundary within several $\lambda_{De}$. As a matter of fact, integrating equation 2. 1. 7 within several $\lambda_{De}$ provides sufficient accuracy. Clearly, one can find in figure 6(b) that increasing emission coefficient mitigates sheath potential. A schematic of space potential variation is given in figure 6(c). Electric field varies linearly inside transition zone and becomes constant outside. Note that surface potential always floats below plasma since $\overline{\Phi_w} > -\Phi_{ext}$.

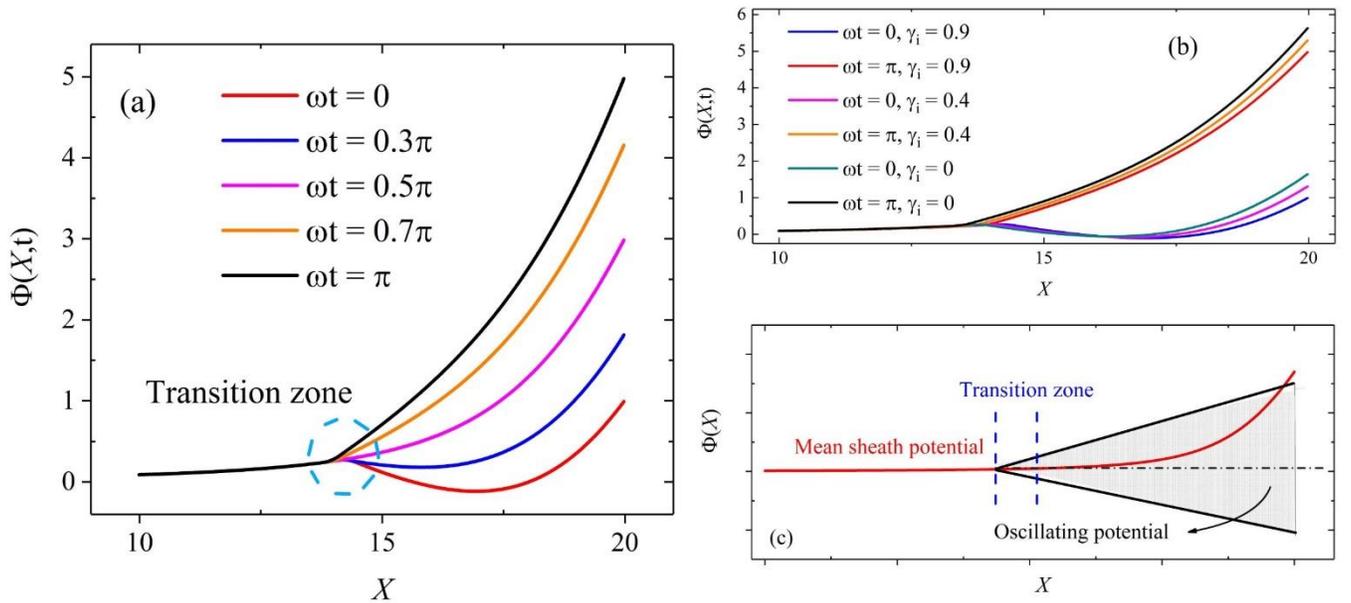

Figure 6. Calculated real-time space potential. (a) Potential at different phases with $\Phi_{ext} = -2$ and $\gamma_i = 0.5$. (b) Comparison of potentials with different emission coefficient. (c) Schematic of space potential profile.

Calculated $\Delta I_{em}$ is compared with $\Phi(X)$ in figure 7. Unsurprisingly, the obtain value of $\Delta I_{em}(X, t)$ is small in contrast to wall potential even at $\omega t = 0$ when space potential is significantly mitigated. This result is echoed by the conclusion we have drawn in the end of section 2.1 that ISEE primarily modifies static wall potential while the presence of secondary electrons barely changes space potential once the boundary condition is fixed. This property greatly simplifies expressions in equation 2. 2. 4-2. 2. 6. In the following deductions concerning sheath capacitance and stochastic heating, the term $\Delta I_{em}$ will be omitted.

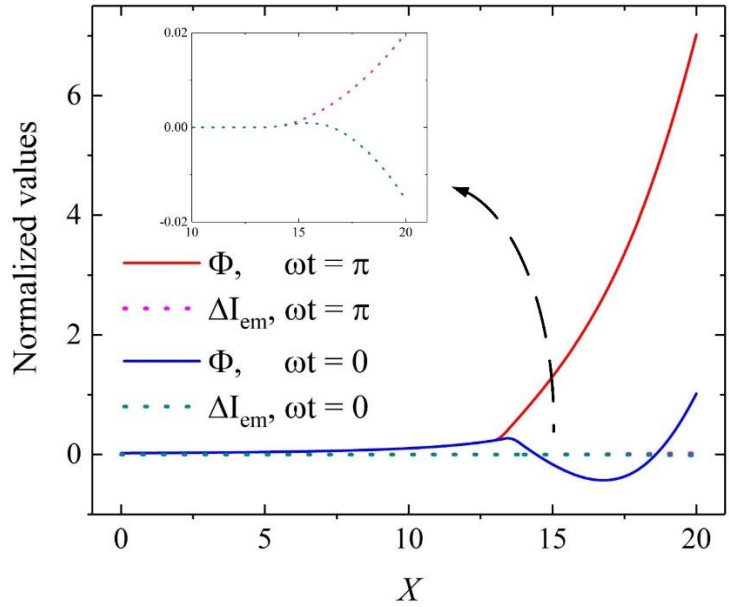

Figure 7. Calculated sheath potential in comparison with corresponding $\Delta I_{em}$ term at different times. Adopted parameters are $\Phi_{ext} = 3$ and $\gamma_i = 0.5$. Zoom-in plot of $\Delta I_{em}$ is shown as well.

Expanding equation 2.2.6 in series, we arrive at:

$$X_{se}(t) = D_{sh} \sum_{n=1}^{\infty} \binom{0.5}{n} \frac{[2\Phi_{ext}\cos(\omega t)]^n}{D_{sh}^{2n}} = -\frac{\Phi_{ext}}{D_{sh}}\cos(\omega t) + \frac{\Phi_{ext}^2}{2D_{sh}^3}\cos^2(\omega t) - \frac{\Phi_{ext}^3}{2D_{sh}^5}\cos^3(\omega t) + \cdots \qquad 2.2.8$$

It is clear that electron sheath front moves sinusoidally if only the first order term is considered. In respect to moderate external voltage where the term $\frac{\Phi_{ext}}{D_{sh}^2} \ll 1$, elements of equation 2.2.8 descend in value at higher orders. Displacement current can be obtained since electric field variation at boundary is known, which is found to be:

$$J_d = \frac{\varepsilon_0 T_{ep}}{e\lambda_{De}}\left[-\frac{\Phi_{ext}}{D_{sh}} + \frac{\Phi_{ext}^2}{D_{sh}^3}\cos(\omega t) - \frac{3\Phi_{ext}^3}{2D_{sh}^5}\cos^2(\omega t) + \cdots\right]\omega\sin(\omega t) \qquad 2.2.9$$

This expression gives a more explicit I-V relation of sheath. One may also directly derive the displacement current using equation 2.2.6, which gives $J_d = -\frac{\varepsilon_0 T_{ep}}{e\lambda_{De}}\Phi_{ext}\,\omega\sin(\omega t)\left[D_{sh}^2 - 2\Phi_{ext}\cos(\omega t)\right]^{-0.5}$. In figure 8 the current waveforms deduced using different expressions are compared. One can find that expansion of 3rd order is very close to the original result, while linear expression is less accurate when value of $\frac{\Phi_{ext}}{D_{sh}^2}$ is greater. If only the linear term is taken, capacitance of a single sheath becomes $C_{sh,single} =$ $\frac{\varepsilon_0 A}{d_{sh}}$ with $A$ the electrode surface, while the total value of two capacitive sheaths connected in series is half the value:

$$C_{sh,tot} = \frac{\varepsilon_0 A}{2d_{sh}} \qquad 2.2.10$$

which is exactly the same with the result from homogeneous model using sinusoidal current as control variable,[5] but is somewhat smaller than the $C_{sh,tot} = 0.613 \frac{\varepsilon_0 A}{d_{sh}}$ given by Lieberman's model.[10] The difference is due to our assumption of uniform ion density. An important conclusion is that the presence of secondary electrons, if no ionization due to SEs is considered, only influences sheath capacitance by changing mean DC sheath size. One can imagine that a similar conclusion can be derived for sheath conductance, which is to be illustrated by analyzing stochastic heating.

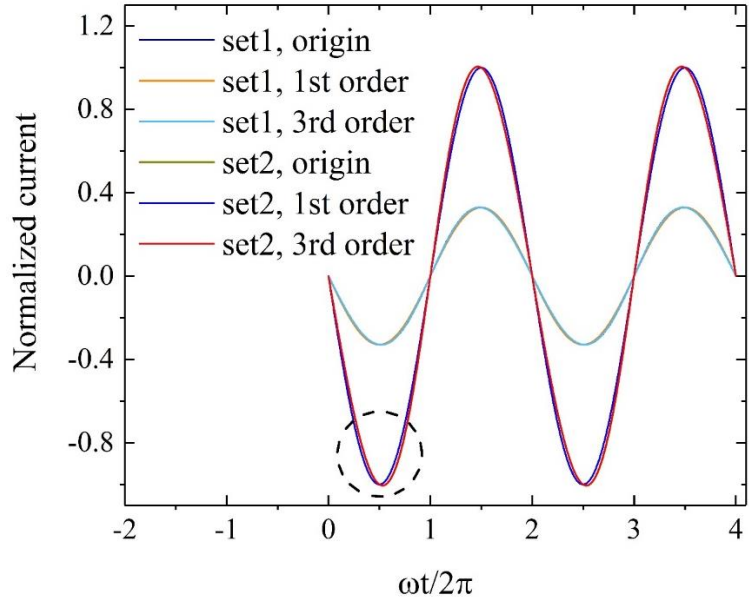

Figure 8. Calculated current waveform. Current is normalized by factor $-\frac{\omega \varepsilon_0 T_{ep}}{e \lambda_{De}}$. In set1 $\Phi_{ext} = -2$, $D_{sh} = 6.1$, in set2 $\Phi_{ext} = -10$, $D_{sh} = 10.0$. Original formula is compared with current expression expanded to 1st and 3rd order. Nonlinearity is more obvious when $\frac{\Phi_{ext}}{D_{sh}^2}$ is greater, one example of nonlinearity is circled out.

Stochastic heating, also referred to as collisionless heating, is caused by the interaction of oscillating sheath with electrons at sheath edge. Distribution of electrons is not uniform, leading to electron density perturbation and absorption of kinetic energy.[12] A hard wall model regards the oscillating plasma sheath as a moving wall which collides with incoming electrons. In hard wall model, power transfer of stochastic heating is given by:[10]

$$S = \int_{v_s}^{\infty} \frac{1}{2} m_e (v_r^2 - v^2)(v - v_s) f_s(x,t) \mathrm{d}v \qquad 2.\,2.\,11$$

where $v_s$ is velocity of electron sheath front, $v_r = -v + 2v_s$ is the velocity after collision and $f_s$ is the EVDF at electron sheath front. Equation 2.2.11 can be interpreted as the energy gained in electron-sheath collision times the number of electron encountering the collision per unit time. After a change of variable in the bulk electrons' frame, it is rewritten as:[10]

$$S = -2m_e \int_0^{\infty} n_s v_s [v'^2 - 2v'(v_s - v_{eb}) + (v_s - v_{eb})^2] g_{eb}(v') \mathrm{d}v' \qquad 2.\,2.\,12$$

with $v' = v - v_{eb}$, $v_{eb}$ the oscillating bulk electron velocity and $g_{eb}(v - v_{eb}) = f_0(v,t)$. Differences with Lieberman's model start from below since we assume a sinusoidal voltage instead of current.

Let conduction current in plasma $J_c$, the current continuity requires that:

$$J_c = e n_0 v_s = e n_{eb} v_{eb} \qquad 2.\,2.\,13$$

Recall that $n_0$ is the electron density at sheath edge. Therefore, $v_s - v_{eb} = \frac{J_c}{e}(\frac{1}{n_0} - \frac{1}{n_{eb}})$. Note that conduction current in plasma equals to displacement current in sheath, provided that conduction current in sheath is too little to be considered. The expression $v_s - v_{eb}$ is thus an odd function in regard to $\omega t$ according to our previous deductions. Consequently, averaging equation 2.2.12 eliminates the first and third terms in RHS, leading to:

$$\langle S \rangle_{\omega t} = 4 m_e \Gamma_{eb} \frac{n_{eb}}{n_s} \langle n_0 v_s (v_s - v_{eb}) \rangle_{\omega t} \qquad 2.\,2.\,14$$

where $\Gamma_{eb} = \frac{1}{4} n_{eb} \langle v_{eb} \rangle = n_{eb} \sqrt{\frac{T_{ep}}{2\pi m_e}}$ is the Maxwellian distribution and $\langle v_{eb} \rangle$ is mean electron speed in bulk plasma. Bringing in the displacement current:

$$\langle S \rangle_{\omega t} = \frac{1}{e^2} \sqrt{\frac{8 T_{ep} m_e}{\pi}} \frac{n_{eb}}{n_0^2} (n_{eb} - n_0) \langle J_d^2 \rangle_{\omega t} \qquad 2.\,2.\,15$$

Recall our postulate of uniform ion density, under which assumption the presheath is flat and bulk electron density is the same as that in electron sheath edge. Therefore, equation 2. 2. 15 inevitably becomes zero in homogeneous model. This is one major drawback of homogeneous model since it predicts no stochastic heating.[38] Fortunately, Kaganovich's previous work suggested a possible correction by implementing two-step ion density model.[11] The plasma-sheath boundary is defined with a sudden change of ion density from $n_{eb}$ in plasma bulk and $n_0$ in sheath. A static potential barrier is supplemented to offset charge discontinuity. Though not very rigorous, [39] the static barrier can be neglected for simplicity, leading to: [40]

$$\langle S \rangle_{\omega t} = \frac{1}{2} m_e \langle v_{eb} \rangle \eta(\eta - 1)(\frac{\varepsilon_0 V_{ext}}{e d_{sh}}\omega)^2 \qquad 2. 2. 16$$

with $\eta = \frac{n_{eb}}{n_0}$. In addition, equation 2. 2. 16 also provides the sheath conductance if we define $\langle S \rangle_{\omega t} = \frac{1}{2} V_{ext}^2 G$:

$$G = m_e \langle v_{eb} \rangle \eta(\eta - 1)(\frac{\varepsilon_0 \omega}{e d_{sh}})^2 \qquad 2. 2. 17$$

Briefly, presented model in section 2.1, 2.2 makes it possible to deduce RF sheath potential analytically in the presence of secondary electron emission. In the collisionless limit, SEE mainly modifies the mean wall potential by virtue of changing flux balance near the boundary. The density of SEs sharply drops down away from the emissive surface, making $N_{em}$ rather low in most areas of the sheath. Consequently, real-time potential in sheath is barely affected by SEE once boundary condition, i.e. mean wall potential is fixed. However, this doesn't mean that sheath capacitance and conductance are independent from surface emission as the size of mean DC sheath is sensitive to $\gamma_i$. In the next section, the implicit relation between sheath capacitance and secondary electron emission will be solved using numerical approach.

## 2.3 Numerical analyses of real-time emissive rf sheath with inhomogeneous model

Involving ion density drop in sheath requires a known mean space potential, which has no analytical expression in Cartesian coordinate as we have shown in section 2.1. Therefore, a numerical solution would be preferable to connect obtained DC sheath with the oscillating component.

A naïve approach is to introduce the notion $\varphi$, representing the phase during which $X_{se}(t) < X$ within half period. In this framework, the mean electron density considering ion density drop becomes: [26]

$$\overline{N_{ep}}(\Phi) = (1 - \frac{\varphi}{\pi}) N_i(\overline{\Phi}) \qquad 2.3.1$$

Setting $\varphi = \omega t$ with $X = X_{se}(t)$ and bringing in results we have derived in section 2.1, the real-time location of sheath edge can be resolved from the equation below:

$$[1 + 2\overline{\Phi}(X_{se})]^{0.5} \exp[-\overline{\Phi}(X_{se})] = 1 - \frac{\omega t}{\pi} \qquad 2.3.2$$

In this manner, the numerical solution of sheath edge is instantly obtained since mean sheath potential $\overline{\Phi}(X)$ is known. However, obtained results is not self-consistent as calculated real-time wall potential does not equal to our prerequisite in equation 2.1.1. The reason is that we underestimate the error brought by the approximation $\overline{N_e}(\Phi) = N_e(\overline{\Phi})$. This approximation has been used to obtain $\overline{\Phi}(X)$ in equation 2.1.19 and is implemented again in equation 2.3.2, which further diverges the results from the accurate sheath solution.

An alternative way is to determine the sheath location directly from Poisson equation, namely finding $X_{se}(t)$ such that:

$$\int_{X_{se}}^{D_{sh}} dX' \int_{X_{se}}^{X'} N_i(X'') dX'' = \overline{\Phi_w} + \Phi_{ext} \cos(\omega t) \qquad 2.3.3$$

Once $X_{se}(t)$ is known, temporal space potential becomes:

$$\Phi(X) = \begin{cases} 0, X < X_{se} \\ \int_{X_{se}}^{X} dX' \int_{X_{se}}^{X'} N_i(X'') dX'', X < X_{se} \end{cases} \qquad 2.3.4$$

Calculation results are given in figure 9. One can compare this with figure 6. General trends of potential in figure 9(a) and (b) are analogous to figure 6(a) and (b), respectively. Yet it is clear that this pure numerical approach avoids the non-monotonic potential profile when $\Phi_{ext} \cos(\omega t) < 0$. Meanwhile, obtained sheath edge oscillates in a broader range than that derived from analytical approach.

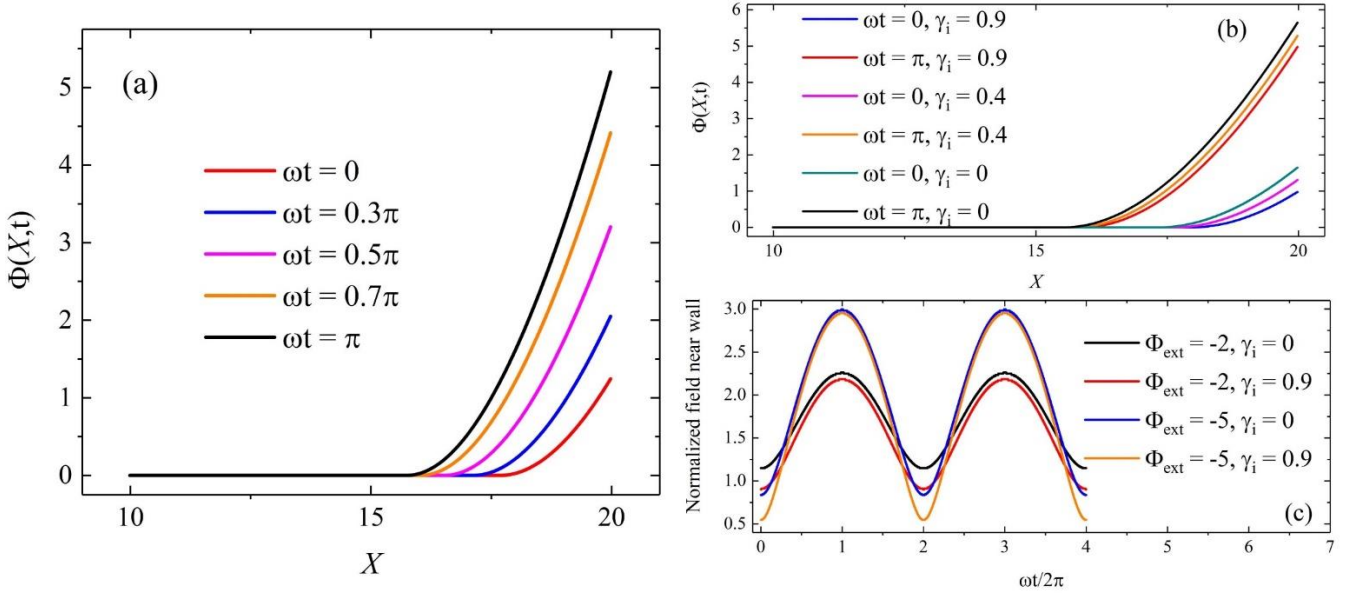

Figure 9. (a) Real-time sheath potential derived from pure numerical method with $\Phi_{ext} = -2$ and $\gamma_i = 0.5$. (b) comparison of sheath potential with different emission coefficients. (c) $\frac{\partial \Phi}{\partial X}$ with different emission coefficients and applied voltages.

This pure numerical approach also allows us to derive sheath capacitance. Since potential of rf sheath is available at any given moment, boundary electric field and consequently displacement current can be calculated as a function of time. By combining like terms, the following relation is deduced for a quick evaluation of sheath capacitance under specific surface emission as well as applied voltage:

$$l = \frac{\rho D_{sh}}{-\Phi_{ext}} \qquad\qquad 2.3.5$$

where $C_{sh,single} = l \frac{\varepsilon_0 A}{d_{sh}}$, and $\rho \sin(\omega t) = \frac{\partial^2 \Phi}{\partial(\omega t)\partial X}$. Note that the waveform of RHS is calculable using oscillating boundary electric field, as is shown in figure 9(c). It Is found that the current is very close to a sinusoidal curve, on account that current-voltage nonlinearity is actually weak as previously discussed in section 2.2. Differentiating figure 9(c) with regard to $\omega t$ gives $l$: 1.77, 1.89, 1.71 and 1.83, corresponding to four sets of parameters in figure 9(c). These values are greater than 1.226 obtained by Lieberman based on sinusoidal current framework as well as value 1 obtained from homogeneous model. The calculated sheath capacitance is influenced by SEE coefficient as we as our choice static sheath edge. Decreasing definition of sheath size or rising up $\delta_{se}$ helps to reduce calculated sheath capacitance.

A flow chart is given in figure 10 to visualize the two sets of methods we introduce to resolve emissive rf sheath. In section 2 we have shown that secondary electron emission mainly manages the mean flux balance. The density of secondary electrons is actually small and their presence barely influences the real-time sheath potential. SEE modifies sheath capacitance and conductance by changing size of mean DC sheath, provided that ionizations induced by secondary electrons are neglected. Before we finish up this section, a simple analysis is given with regard to power loss on electrodes in low collisionality regime.

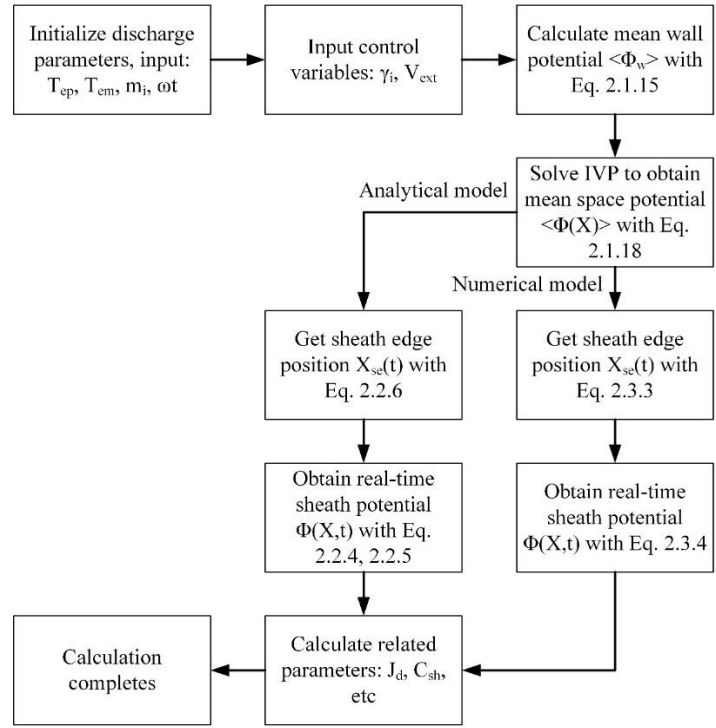

Figure 10. Schematic of two approaches to solve emissive rf sheath.

The power loss of ion on electrode without SEE is the product of ion current with mean DC voltage:

$$P_{ion} = 2en_i\sqrt{\frac{T_{ep}}{m_i}}\overline{\Phi_w} \qquad\qquad 2.3.6$$

Knowing that all secondary electrons finally get lost at electrodes, the power loss of both SE and ion is:

$$P_{ion} + P_{SEE} = 2(1+\gamma_i)en_i\sqrt{\frac{T_{ep}}{m_i}}\{\ln\left(\frac{1}{1+\gamma_i}\sqrt{\frac{\mu}{2\pi}}\right) + \ln[I_0(-\Phi_{ext}) - \frac{\gamma_i}{2\pi\sqrt{2\mu}}H]\} \qquad 2.3.7$$

which has no analytical expression due to term $H$. We define the normalized power loss in respect to the case without SEE as follows:

$$\tilde{P} = \frac{P_{ion} + P_{SEE}}{P_{ion,no\ SEE}} = (1 + \gamma_i) \frac{\overline{\Phi_w}}{\overline{\Phi_w}|_{\gamma_i=0}} \qquad\qquad 2.3.8$$

See equation 2. 1. 13, 2. 1. 16 for expressions of wall potential. The calculation results are given in figure 11. Obviously $\tilde{P}$ climbs up with $\gamma_i$. For very high applied voltage the wall potential is hardly changed by emission coefficient thus it presents a linear relation. The floating sheath limit is a known function

$$\tilde{P}|_{\Phi_{ext}=0} = (1 + \gamma_i) \frac{\ln(\frac{1}{1+\gamma_i}\sqrt{\frac{\mu}{2\pi}})}{\ln(\sqrt{\frac{\mu}{2\pi}})}.$$ Note that the ion density is assumed to be constant. In realistic PIC

simulation, the density is formed self-consistently under specific pressure thus ionization occurs even when background pressure is low (around decades mTorr). In this condition, the ion density is no longer constant under changing emission coefficient. A description of SE's ionization will be given in the next section.

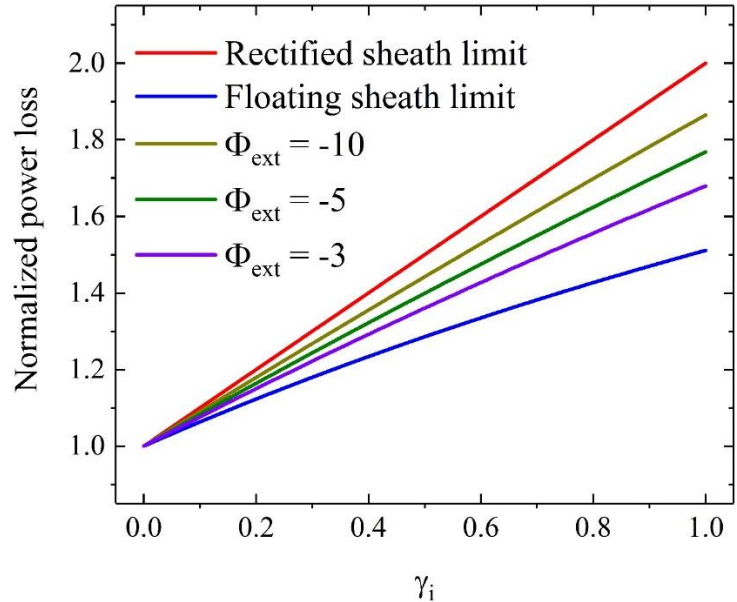

Figure 11. Obtained power loss due to ion and SEE. In rectified limit $\Phi_{ext} = -\infty$, in floating limit $\Phi_{ext} = 0$.

# 3 Discussions on Obtained Results

To validate results obtained in section 2 and justify adopted assumptions, we perform simulation and compare theoretical predictions with simulation results. The first implemented simulation program is based on a continuum kinetic code. The simulation produces noise-free data and clear sheath structure of a plasma bounded by two emissive surfaces. The emission coefficients are set to be equal at both boundaries, and voltage is applied as Dirichlet boundary condition at two surfaces. Note that this 1D1V kinetic simulation neglects ionization and keeps bulk plasma density as constant by using source operator to offset wall losses. It solves 1D kinetic-Poisson equation and advances the velocity distribution function. Collision is characterized by collision operator without emulating real particles. Thus obtained results are irrespective of background pressure and corresponds better to theoretical predictions, c.f. particle-in-cell simulation. More detailed descriptions on simulation methods are given in references. [26,33] PIC simulation results are also provided in the end to illustrate the influence of ionization.

Obtained real-time space potential is given in figure 12. One can compare space potentials in figure 12 with figure 6(a) and figure 9(a). Clearly the numerical model is closer to the simulation results. Recall that the analytical model presents a non-monotonic potential profile which was also observed in previous work without secondary electron emission. [14] The reason of this seemingly contradictory feature is due to assumption of electron density profile. Note that in analytical model charge density in presheath-sheath edge is chosen as constant, so the density drop is within transition zone is ignored, leading to non-monotonic sheath potential. Numerical model considers such density drop so obtained results are closer to the simulation.

Additionally, the electron density drops smoothly in sheath and there is no clear-cut electron sheath front moving back and forth around static ion sheath. This explains the difference between results of numerical model and simulation. Also, there appears some oscillations on plasma bulk potential, which is possible because a true equilibrium can never be achieved and conduction current in bulk plasma contributes to a variation of potential. This can be explained by current continuity, with displacement current in sheath being equal to conduction current of bulk plasma.

Briefly, both analytical and numerical models have pros and cons. The former allows explicit calculation of discharge parameters though theory predictions are not rigorous. The latter provides a better precision but requires computational resources. Their applications in practice should depend on the exact condition where a CCP model is in need.

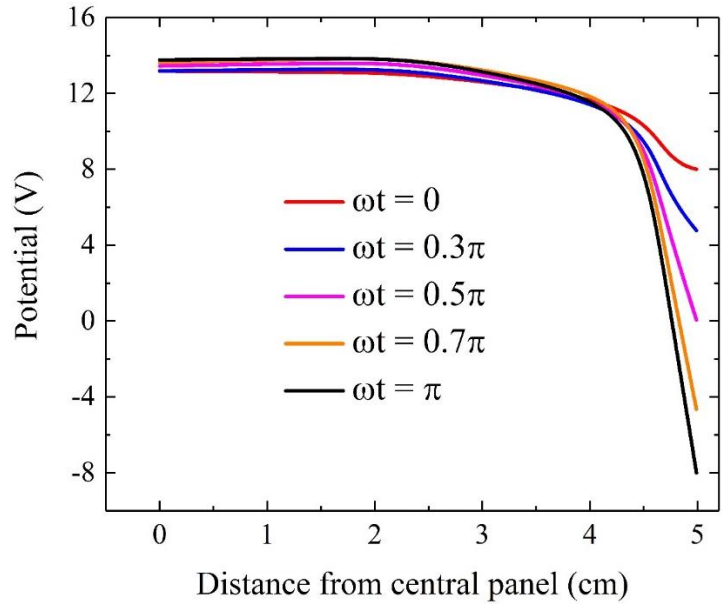

Figure 12. Real-time space potential in sheath from kinetic simulation. We set $T_e = 4$ eV, $T_i = 0.1$ eV $\Phi_{ext} = -2$ and $\gamma_i = 0.5$. Obtained $\overline{\Phi_w} = 3.3$. Note that potential is 0 at electrode when $\omega t = 0.5\pi$ in simulation while in theory part we choose sheath edge potential as 0.

VDFs of electron and ion are shown in figure 13. The VDFs are recorded when left electrode potential is minimum and potential on the right side is maximum. One can clearly observe a region depleted of electrons in figure 13(a) on the left side as potential barrier is maximum for the left plasma sheath. On the right side the sheath contains more electrons since potential barrier is smaller. Surface emissions on both sides are circled out. SE's density on the left quickly vanishes away from surface and SEs are greatly accelerated by sheath potential. On the contrary, emission on the right side is more extensive and is partially merged with plasma electrons. Above characteristics are consistent with the schematic in figure 5 with right side the case $\omega t = 0$ and left side $\omega t = \pi$. IVDF is given in figure 13(b), while the ion distribution is almost unchanged within a period since ions response to mean space potential.

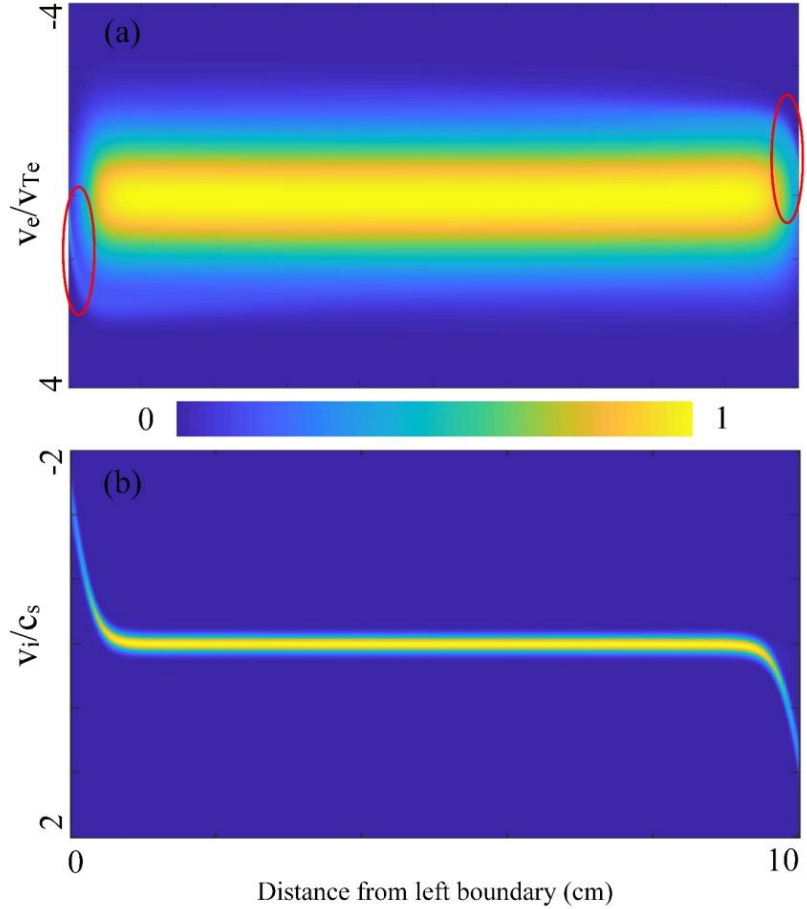

Figure 13. (a) Normalized EVDF when voltage at left electrode is minimum, velocity axis is normalized in regard to thermal speed $v_{Te} = \sqrt{\frac{T_{ep}}{m_e}}$. Surface emissions at both boundaries are marked out with circle, and a region depleted of electrons appears near left boundary. EVDF is asymmetrical in space. (b) Normalized IVDF at same time as (a), velocity axis is normalized in regard to sound speed $c_s = \sqrt{\frac{T_{ep}}{m_i}}$. Results are recorded before equilibrium to make surface emission obvious. Ions are accelerated towards the surfaces and IVDF is symmetrical since ions respond to time-averaged potential.

Temporal voltage and electric field at boundary are sketched in figure 14. After around 1 µs waveform of wall electric field is stabilized. Obtained data make it possible to derive capacitance of the single sheath. We first calculate the sheath capacitance from simulation data and determine the sheath size using two approaches, leading to $C_{sh,single} = 2.05 \frac{\varepsilon_0 A}{d_{sh}}$ using sheath edge definition $\delta_{se} = 5\%$, and $C_{sh,single} = 2.73 \frac{\varepsilon_0 A}{d_{sh}}$ using Bohm criterion. Obtained results are considerably greater than the homogeneous model in section 2.2 and is closer to results using inhomogeneous model in section 2.3. It seems that using Bohm criterion to judge sheath edge is less accurate than the method we introduce, i.e. defining a significant breakdown of charge neutrality with $\delta_{se} = \frac{n_i - n_e}{n_i}$.

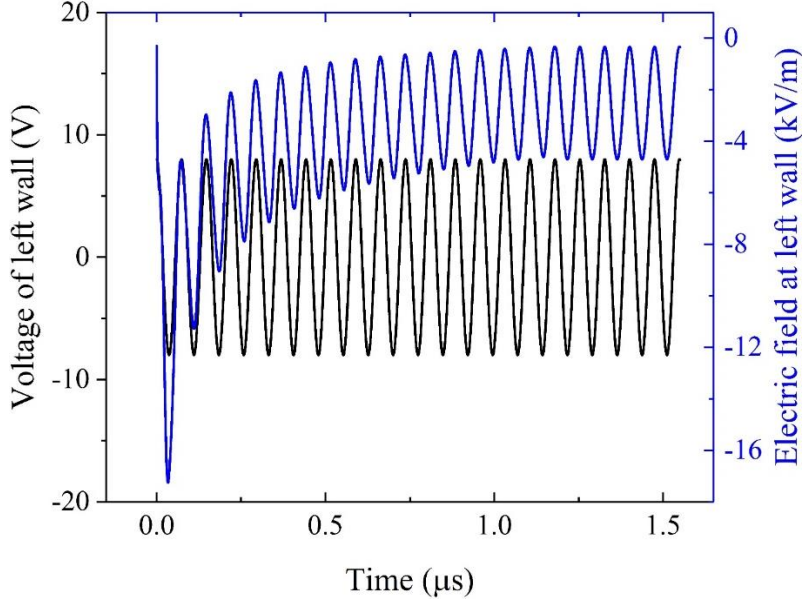

Figure 14. Voltage and electric field at left wall from simulation. Electric field (also displacement current) becomes stable around 1 μs. Sheath capacitance is calculated to be $C_{sh,single} = 2.05 \frac{\varepsilon_0 A}{d_{sh}}$ using sheath edge definition $\delta_{se} = 5\%$, and $C_{sh,single} = 2.73 \frac{\varepsilon_0 A}{d_{sh}}$ using Bohm criterion.

In a more realistic scenario, the plasma density cannot be kept constant with changing emission coefficient due to ionization of secondary electrons. In addition, most particle-in-cell simulations assign an initial plasma density and let the system develop self-consistently. So stabilized plasma parameters like electron temperature, plasma density cannot be directly controlled. We implement PIC simulation with low background pressure and changing emission coefficients. The PIC code incorporates adaptive particle management and parallel computing, providing accurate real-time CCP discharge dynamics. Some simulation parameters are given here: length 6.7 cm, gas type hellion with temperature 300K, applied voltage 210V and frequency 13.56 MHz, pressure 4-8 Pa. Electron-neutral collision (elastic collisions, excitation and ionization), the ion-neutral gas collision, ion-induce secondary electron emission are taken into account. More details of code algorithm are presented in reference.[41, 42]

Space ion density from simulation is shown in figure 15. Clearly ion density in panel center, or in other words plasma bulk, rises up with emission coefficient $\gamma_i$, as is shown in figure 15(a). In figure 15 (b) relation between bulk plasma density and $\gamma_i$ under different pressures is compared, it can be seen that under lower pressure bulk plasma density is collinear with emission coefficients while non-linearity appears for higher pressure. In theories presented in section 2, charge densities are normalized so change of plasma density does not alter obtained results, provided that collisionality in sheath is low.

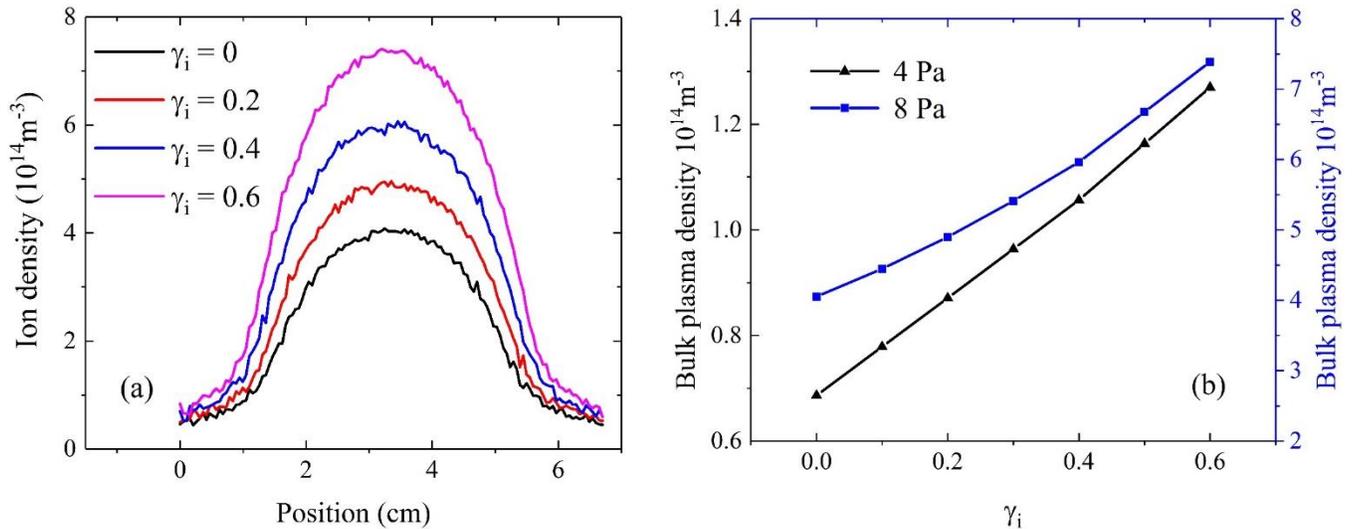

Figure 15. (a) Ion distribution in space with different emission coefficients in 8 Pa. (b) Stable bulk plasma density with different emission coefficients and pressures.

Apart from ionization, the excitation also plays a role in discharge dynamics. In low pressure regime, one can imagine that electron energy is slightly reduced due to inelastic collisions in excitation. At higher pressure like atmospheric pressure, CCP can become Penning-dominated. [43,44] In this regard, Penning ionization due to metastable atoms is crucial and excitation becomes more influencial than in low-pressure regime.

Remarkably, above theories are irrespective of electron temperature in bulk plasma. In reality, plasma electron temperature is determined self-consistently from power balance of discharge system. In PIC simulation, it is possible to track the stabilized space potential under given pressure, as is shown in figure 16(a). The plasma electron temperatures under different $\gamma_i$ and pressures are given in figure 16(b). It is clear that sheath potential decreases when increasing emission coefficient. This is consistent with results in figure 1.

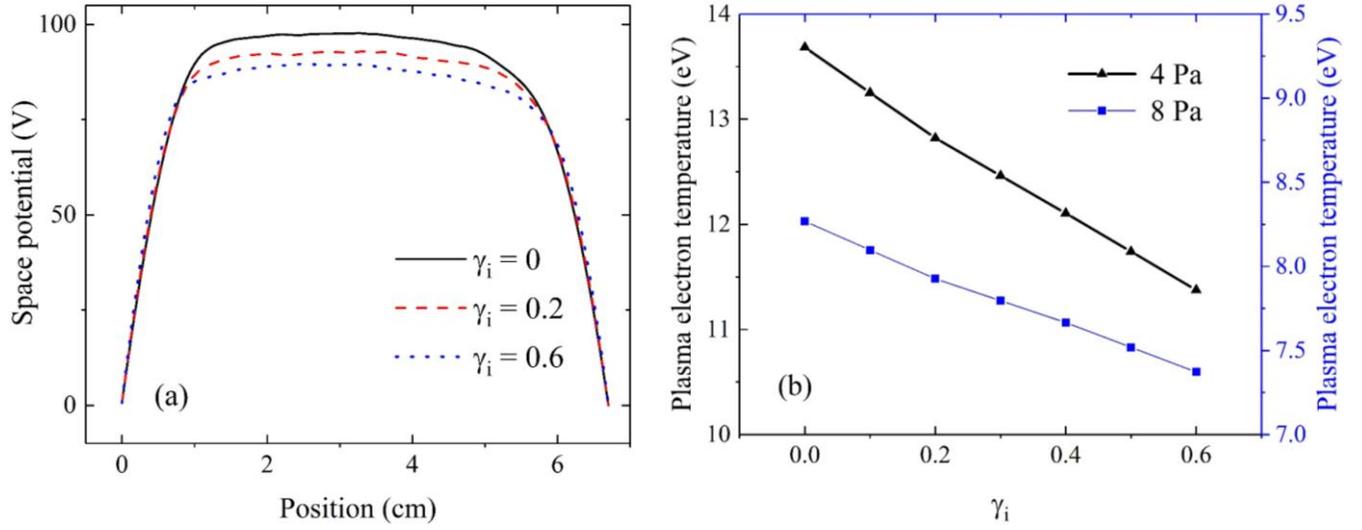

Figure 16. (a) Space potential with different emission coefficients in 8 Pa. (b) Plasma electron temperature with different emission coefficients and pressures.

However, the increase of plasma density, either due to higher pressure or emission coefficient, inevitably modifies collisionality in both bulk plasma and rf sheath. Generally, ionization rate increases with emission coefficients. [42] Thus a more rigorous theory involving weak collisionality is more preferable for higher pressure. Continuously augmenting background pressure induces mode transition and CCP discharge becomes dominated by ionizations of SEs from surface, which is beyond the scope of this paper but is expected to be expatiated in future works.

## 4 Conclusion

In this work we analyze the role of ion-induced secondary electron emission in low pressure capacitively-coupled radio-frequency sheath. A theoretical ground is established which is able to predict mean wall potential relative to bulk plasma as a function of secondary emission coefficient as well as applied voltage. It is found that SEE mitigates mean wall potential but the influence is smaller at higher voltage amplitude. Average potential in sheath is completely resolved and the real-time space potential is calculated with an analytical model as well as a numerical model, respectively. The homogeneous model using known sinusoidal voltage yields the same sheath capacitance with that obtained with known sinusoidal current, while inhomogeneous model provides more precise solution. SEE influences sheath capacitance by changing size of mean DC sheath. Sheath conductance is calculated and is dependent on ratio of ion density at plasma bulk over density at plasma sheath edge. Sheath conductance is influenced by SEE in a similar manner as sheath capacitance. Derived results are justified by a kinetic simulation which neglects ionizations in space. Using a full PIC simulation, we find that collisionality in sheath rises up with stronger SEE, making collisionless assumption inapplicable. A self-consistent theory under higher collisionality involving ionization of secondary electrons is expected in further works.

**ACKNOWLEDGMENTS**

Authors thank Dr. M. D. Campanell for his help in constructing simulation model and Dr. I. D. Kaganovich for fruitful discussions. This research was partly supported by National Natural Science Foundation of China (Grant No. 51827809, U1830129, 51707148)